\documentclass[sigconf]{acmart}
\AtBeginDocument{%
  \providecommand\BibTeX{{%
    \normalfont B\kern-0.5em{\scshape i\kern-0.25em b}\kern-0.8em\TeX}}}

\usepackage{subcaption}
\copyrightyear{2024}
\acmYear{2024}
\setcopyright{acmlicensed}\acmConference[IMX '24]{ACM International Conference on Interactive Media Experiences}{June 12--14, 2024}{Stockholm, Sweden}
\acmBooktitle{ACM International Conference on Interactive Media Experiences (IMX '24), June 12--14, 2024, Stockholm, Sweden}
\acmDOI{10.1145/3639701.3656308}
\acmISBN{979-8-4007-0503-8/24/06}
\begin{document}

\title{Human Interest or Conflict? Leveraging LLMs for Automated Framing Analysis in TV Shows}

\author{David Alonso del Barrio}
\email{ddbarrio@idiap.ch}
\affiliation{%
  \institution{Idiap Research Institute}
  \country{Switzerland}
}
\author{Max Tiel}
\email{mtiel@beeldengeluid.nl}
\affiliation{%
  \institution{Nederlands Instituut voor Beeld en Geluid}
  \country{The Netherlands}
}
\author{Daniel Gatica-Perez}
\email{gatica@idiap.ch}
\affiliation{%
  \institution{Idiap Research Institute and EPFL}
  \country{Switzerland}
}


\begin{abstract}
%ORIGINAL VERSION BY DGP
\footnote{\textbf{David Alonso del Barrio, Max Tiel, Daniel Gatica-Perez| ACM 2024. This is the author's version of the work. It is posted here for your personal use. Not for redistribution. The definitive Version of Record will be published in IMX'24 ACM International Conference on Interactive Media Experiences https://doi.org/10.1145/3639701.3656308.}}
In the current media landscape, understanding the framing of information is crucial for critical consumption and informed decision making. Framing analysis is a valuable tool for identifying the underlying perspectives used to present information, and has been applied to a variety of media formats, including television programs. However, manual analysis of framing can be time-consuming and labor-intensive. This is where large language models (LLMs) can play a key role. 
In this paper, we propose a novel approach to use prompt-engineering to identify the framing of spoken content in television programs. Our findings indicate that prompt-engineering LLMs can be used as a support tool to identify frames, with agreement rates between human and machine reaching up to 43\%. As LLMs are still under development, we believe that our approach has the potential to be refined and further improved. The potential of this technology for interactive media applications is vast, including the development of support tools for journalists, educational resources for students of journalism learning about framing and related concepts, and interactive media experiences for audiences.

%ORIGINAL VERSION BY DAVID
%In the current media landscape, understanding the framing of information is crucial for critical consumption and informed decision making. Framing analysis is a valuable tool for identifying the underlying perspectives used to present information, and has been applied to a variety of media formats, including television programs. However, manual analysis of framing can be time-consuming and labor-intensive. This is where large language models (LLMs) can play a key role. 
%In this paper, we propose a novel approach to use prompt-engineering to identify the framing of spoken content in television programs. Our findings indicate that prompt-engineering LLMs can significantly be a support tool identifying frames, with agreement rates between human and machine reaching up to 43\%. While this is still under development, we believe that our approach has the potential to be refined and further improved.
%The potential applications of this technology are vast, including the development of support tools for journalists, educational resources for students of journalism to understand the concept of framing, and interactive media experiences for viewers.

\end{abstract}

\begin{CCSXML}
<ccs2012>
   <concept>
       <concept_id>10003120.10003121.10011748</concept_id>
       <concept_desc>Human-centered computing~Empirical studies in HCI</concept_desc>
       <concept_significance>500</concept_significance>
       </concept>
   <concept>
       <concept_id>10010147.10010178.10010179.10003352</concept_id>
       <concept_desc>Computing methodologies~Information extraction</concept_desc>
       <concept_significance>500</concept_significance>
       </concept>
 </ccs2012>

\end{CCSXML}

\ccsdesc[500]{Human-centered computing~Empirical studies in HCI}
\ccsdesc[500]{Computing methodologies~Information extraction}

\keywords{media, TV, LLMs, prompt-engineering, framing analysis }


\maketitle
\section{Introduction}

%revised version
% frame analysis and its importance in understanding media content.
Framing analysis is a technique for understanding how media content shapes our perception of the world, as it helps media audiences identify the underlying perspectives that are used to present information \cite{de2005news}. Identifying frames in news stories allows to understand what aspects of the story are being emphasized. For example, given two news items about the same topic such as the war between Russia and Ukraine, one may emphasize the economic repercussions in Europe due to this event (economic frame), while the other may show the personal case of a family displaced by the war and left homeless (human interest frame). Understanding these frames can support media readers and viewers to act as more critical consumers of media and to make informed decisions about the consumed information  \cite{aalberg2012media}.

%original version
% frame analysis and its importance in understanding media content.
%Framing analysis is a valuable tool for understanding how media content shapes our perception of the world. It helps us identify the underlying perspectives that are used to present information \cite{de2005news}. Understanding these frames enables us to be more critical consumers of media and to make informed decisions about the information we consume \cite{aalberg2012media}.

% AI AND JOURNALISM
% IMX papers
In recent years there have been great advances in the combination of journalism and computing. Concepts closely related to interactivity such as Automated Journalism are emerging \cite{diakopoulos2019automating, sina}, further emphasised by the latest developments in artificial intelligence \cite{alfonso, kim, soe}.

% LLMS
The development of  Large Language Models (LLMs) has opened up new possibilities for media content analysis \cite{chew2023llm}. LLMs are large neural networks trained on large amounts of text data. Recently, they have proven to be effective in a variety of domains, including journalism \cite{fiona, Hadi_2023} and natural language processing tasks, such as text classification, machine translation or summarization \cite{10367969, yao2023empowering, zhang2023benchmarking}.
% LLMS and frame analysis
The use of Large Language Models (LLMs) for frame analysis is a relatively new field of research, but it is attracting increasing interest as LLMs are effective in identifying frames in text and can be applied to a variety of media formats, such as headlines, tweets, or news articles \cite{alonsoframing, Gilardi_2023, alizadeh2023opensource}. This has ignited a growing interest in their potential application to spoken content, such as television transcripts, which is the focal point of our work.
Framing analysis plays a pivotal role in comprehending how news programs shape our perception of reality. By identifying the underlying frames employed in newscasts, we can gain insights into the perspectives presented and the potential influence on public opinion \cite{entman2009nature}. However, conventional framing analysis methods are often tedious and labor-intensive \cite{bhatia2021openframing}, thus restricting their applicability to large-scale analysis. LLMs offer a viable solution to this challenge by automating the frame detection process, enabling more efficient and comprehensive analysis of media content.

To address this research gap, we pose three research questions:

%revised
{\bf RQ1}: How accurately can a GPT model identify frames in television programs using a prompt-engineering approach?
%original
%{\bf RQ1}: How accurate can a GPT model is the GPT-3.5 model at identifying frames in TV show transcripts compared to human annotators?

{\bf RQ2}: In comparison to human annotation, what are the limitations of a GPT-3.5 model in classifying frames of TV show transcripts?
%revised
%{\bf RQ2}: In comparison to a human annotator, how well does the GPT-3.5 model perform in classifying frames of TV show transcripts?

{\bf RQ3}: How can GPT models and prompt engineering be used in potential interactive applications of framing analysis?

%original
%{\bf RQ2}: What are the potential applications of LLM-based framing analysis?

%revised
Our research addresses these questions by making the following contributions:

1. We developed and validated a framework that leverages LLMs for automated frame detection in TV transcripts. A dataset has been created with transcripts of two current affairs programs on Dutch television. These program extracts have been classified by GPT-3.5 using a prompt composed by the definition of different frame types, the transcription item and finally we asked the model to choose the predominant frame based on the definitions.
% {\bf @David}; We would have to expand this part to give details of what we actually did (which so far is missing in the introduction.) Collected new data, manually annotated with experts, generated prompt-engineering frames using GPT-3.5, compared human vs. machine,

2. We  created an annotation system based on Google Forms where an expert read the transcription to be annotated as well as the frame definitions, and answered a series of questions that justified their answer. After that annotation, we analyzed, together with experts in media and frame analysis, the results obtained from the agreement/disagreement between human and machine.

3. We explored potential integration strategies to incorporate LLMs into existing media workflows, empowering journalists, media consumers, and researchers towards improved and better informed media analyses.
%original
%Our research addresses these inquiries by developing a framework that leverages LLMs for automated frame detection in TV transcripts. We also explore potential integration strategies to incorporate LLMs into existing media workflows, empowering journalists, media consumers, and researchers to harness the power of LLMs for more effective and informed media analysis.
%revised
Our research could support journalists to expeditiously verify (or identify) the framing employed in their news articles as conveyed by their text. This capability can significantly enhance accuracy and objectivity in their reporting. Additionally, our research empowers media consumers to gain a deeper understanding of the perspectives presented in TV news, by fostering critical consumption and informed decision-making.
Our research represents a step forward in automating frame detection and integrating LLMs into media workflows. This has the potential for more comprehensive and insightful media analysis, ultimately enriching the media landscape for all stakeholders.
%original
%One of the key contributions of our research lies in enabling journalists to expeditiously identify the framing employed in their news articles. This capability can significantly enhance accuracy and objectivity in their reporting. Additionally, our research empowers media consumers to gain a deeper understanding of the perspectives presented in TV news. This fosters critical consumption and informed decision-making.
%We firmly believe that our research represents a significant step forward in automating frame detection and integrating LLMs into media workflows. This paves the way for more comprehensive and insightful media analysis, ultimately enriching the media landscape for all stakeholders.

The paper is organized as follows. In Section \ref{section:related_work}, we discuss related work. In Section \ref{section:data}, we describe the news dataset. In Section \ref{section:methodology}, we describe the methodology for both human labeling and machine classification of news frames. We present the results and discuss them in Section \ref{section:results}.
%and we discuss them in Section \ref{section:discussion}.
Finally, we provide  conclusions in Section \ref{section:conclusions}.

\section{Related Work}\label{section:related_work}
% Review existing research on framing analysis in media studies and journalism.
As proposed by Entman, "to frame is to select some aspects of a perceived reality and make them more salient in a communicating text, in such a way as to promote definition, causal interpretation, moral evaluation, and/or treatment recommendation for the item described" \cite{entman1993framing}. This definition was the starting point for a long history of research on framing analysis in journalism and media studies \cite{de2005news}. 
There are two main approaches to framing analysis: an inductive one, in which frames emerge from the text \cite{van2012frames}, and a deductive one, in which, given already defined types of frames, they are tried to be identified in the text \cite{dirikx2010frame}. There is also a second sub-classification. There are generic frames \cite{semetko} \cite{dallas} such as economic, conflict, etc., and issue-specific frames such as frames about Covid \cite{wicke2020framing} or climate change \cite{badullovich2020framing}. At the same time, frames have been studied in different types of content, whether in the written press or on television \cite{claes}.

In terms of computing research in this field \cite{ali2022survey}, including attempts to automate the task,  much of the previous work was done 
%much of the previous work has been done as a part of this concept of 
in the domain of Computational Journalism \cite{10.1145/2001269.2001288} \cite{ali2022survey}.
Dallas et al. \cite{dallas} created a dataset with articles about polemic topics such us immigration, same-sex marriage, or smoking, and they defined 15 types of frames; for each article, annotators were asked to identify any of the 15 framing dimensions present in the article, and to label spans of text which cued them, based on the definitions of each of the frame dimensions, while stating the main frame of each article.
This dataset has been used in the work of Khanehzar et al. \cite{Khanehzar2019ModelingPF}, where they used conventional classification techniques like Support Vector Machines (SVM) as a baseline method 
and showcased the enhancement in frame categorization by employing pre-trained language models like Bert, ROBERTa, and XLNet, through a fine-tuning methodology. This dataset was recently used for a task about detection of frames in SemEval-23, an International Workshop on Semantic Evaluation \cite{piskorski2023semeval}.

With the various advances in Natural Language Processing, the previous literature shows how new technologies have been adapted to the way frames are identified. This spans the use of corpus linguistics software to extract frames by  Touri et al. \cite{touri2015using}; the use of topic modeling techniques by Walter et al. \cite{walter2019news}, that describes computational methods to inductively analyse framing in texts; %but in an inductive through topic modeling ;  
and the use of fine-tuned BERT based models \cite{liu2019detecting}.
Recently, there is a growing number of studies on the use of LLMs in Data Journalism and Media \cite{bahare}. Fatemi et al. \cite{10367969} explored the potential of GPT in a zero-shot setting for multi-class classification of news articles. Bianchini et al. \cite{bianchini2023using} attempt to classify various dimensions of freedom (which are defined in the prompt) in several interviews with Peruvian political leaders using OpenAI models to do the transcription and then the classification of abstract concepts like freedom, which often poses challenges to computational analysis. 
More specific to our paper (using LLMs for framing detection), we can highlighted the works of Gilardi et al. \cite{Gilardi_2023} and Alizadeh et al. \cite{alizadeh2023opensource}, where they classified the 15 frame types of Dallas et al. \cite{dallas} in tweets and news articles, using a prompt-engineering approach that first gives the definitions of the frames and the text to classify, and then asks the model about the most predominant frame.  

Our work follows a deductive approach, i.e., we use a set of frames already defined in the literature. These frames are generic and our content are transcripts of television programs. Specifically, the generic frames are those presented by Semetko et al. \cite{semetko} which defines 5 types of frames: human interest, conflict, economic, morality, and attribution of responsibility. In their original work, they classified both print and TV content, but only manually. This classification system was recently used by Alonso del Barrio et al. \cite{alonsoframing}, who also used LLMs, but in this case their dataset was headlines about Covid vaccination.

% INTERACTIVITY WITH LLMs
Our research has the potential to 
%improve the quality of media content by helping 
support journalists in their daily tasks and help the public become more critical consumers of information \cite{gorkovenko}. 
There are several works on the use of LLMs to create interactive tools  \cite{10.1145/3532106.3533533} in different domains \cite{bang2023multitask}. Our paper moves in the direction of supporting interactivity and the use of LLMs in journalism, not just as a text generator but as an analytical tool.

\section{Data}\label{section:data}
In this section, we describe the dataset and pre-processing steps.

\subsection{Dutch TV news dataset} 
The dataset we used is a selection of 2000 news media items from broadcasts of the public Dutch television news programs EenVandaag (1000 items) and Nieuwsuur (1000 items). 

EenVandaag \footnote{https://eenvandaag.avrotros.nl/} (OneToday) is a daily evening program broadcasted on Dutch public television channel NPO1. EenVandaag has the format of a news program with current issues %actualities 
and background information behind the news. The program is about 30 minutes long and deals with various news topics during an episode. The program has multiple presenters introducing various news items, and also interview experts live in the studio. 

Nieuwsuur \footnote{https://nos.nl/nieuwsuur} (News Hour) is also an evening program and is broadcasted on NPO2, a Dutch public television channel. The broadcasts are between 30 and 45 minutes long, and also have the format of a news program with current issues %actualities 
and background information behind the news. This program also has multiple presenters and live interviews with experts. 

We chose these two current newscasts as they provide a good overview of Dutch news items on a daily basis. These shows also provided a large corpus of items over the years, with  little change in the show format. Due to this, the data is very consistent. 

For analysis, the spoken words in the video recordings of these programs are transcribed. This was done with the open-source, Kaldi automatic transcriber  \cite{Povey_ASRU2011}. This software can automatically transcribe Dutch spoken language into text. This pre-processing step resulted in a dataset of 2000 texts covering news between 2014 and 2018, varying in length, with an average number of 499 words for Nieuwsuur and 664 words for EenVandaag. As these texts are automatically transcribed and are thus a literal transcription of the spoken words, they contain errors at word level and also regarding sentence construction. The noisy transcriptions 
%As far as we have noticed, this mainly 
caused some issues for the human annotator (Section \ref{section:annotation}), as it involved more time to analyze some of the texts.
% In the original dataset there were these columns: program\_id, scene\_id, carrier\_id, start, end, title, running\_order, date, text\_dutch	
%499.49 
%664.77 

\subsection{Translation of content}

To be able to use GPT-3.5 in an optimal way, and to obtain results in which the language would not be detrimental to model performance, we translated the content into English using the Deepl API \footnote{https://www.deepl.com/es/pro-api?cta=header-pro-api/}. We created a script that called the API; for each entry in Dutch, it translated the text; the English translation is the text used for further analuysis. As mentioned earlier in this section, the original Dutch content contained some noise, in the sense that automatically transcribed text is not entirely clean on some occasions; this noise has been transferred to the translation. The average number of words in the English content corresponds to 496 for Nieuwsuur and 650 for EenVandaag.

%496.17 650.64 

\section{Methodology}\label{section:methodology}

In this section, we explain our methodology for framing analysis, first describing the annotation process, and then summarizing the prompt-engineering technique. 

%Once the working dataset was built, we started to work with it, from the human side with the annotation, as well as from the technical side with the use of prompt-engineering technique. 

\subsection{Annotation} \label{section:annotation}
For this task, we engaged a person with university-level education and a basic knowledge in the field of framing. This was needed to make sure that the annotator was aware of the potential for human bias and was thus able to look at the texts in the most objective way possible. More specifically, the annotator had a background in language and cultural studies. 

%For this task, we engaged a person who had an academic level of thinking and a basic knowledge in the field of framing. This is needed to make sure that the annotator is aware of the risk of human bias and is thus able to look at the texts in the most objective way possible. The annotator had a background in language and culture studies. 

To carry out the labeling of the frames in the data, we first designed a codebook with several definitions of the frame analysis concept. This was then used to train the annotator on the task.
%explain the goal of the task that the annotator had to do. 
In addition, we created an interactive environment through Google Forms to do the annotation. Using Google Apps Script \footnote{https://www.google.com/script/start/}, we created a script that allowed us to generate forms automatically. 

In the annotation form, we first showed the piece of text to annotate, followed by the definitions of the 5 types of frame proposed by \cite{semetko}:
%In the annotation form script, we defined the output structure of the form. First we take the piece of text to annotate, followed by the definitions of the 5 types of frame proposed by \cite{semetko}:

\begin{itemize}
    \item Attribution of responsibility. This frame presents an issue or problem in such a way as to attribute responsibility for its cause or solution to either the government or to an individual or group 
\item Human interest. This frame brings a human face or an emotional angle to the presentation of an event, issue, or problem. 
\item Conflict. This frame emphasizes conflict between individuals, groups, or institutions as a means of capturing audience interest.
\item Morality. This frame puts the event, problem, or issue in the context of religious tenets or moral prescriptions.
\item Economic. This frame reports an event, problem, or issue in terms of the consequences it will have economically on an individual, group, institution, region, or country.
\end{itemize}

\begin{figure}[h]
  \centering
\includegraphics[width=\linewidth]{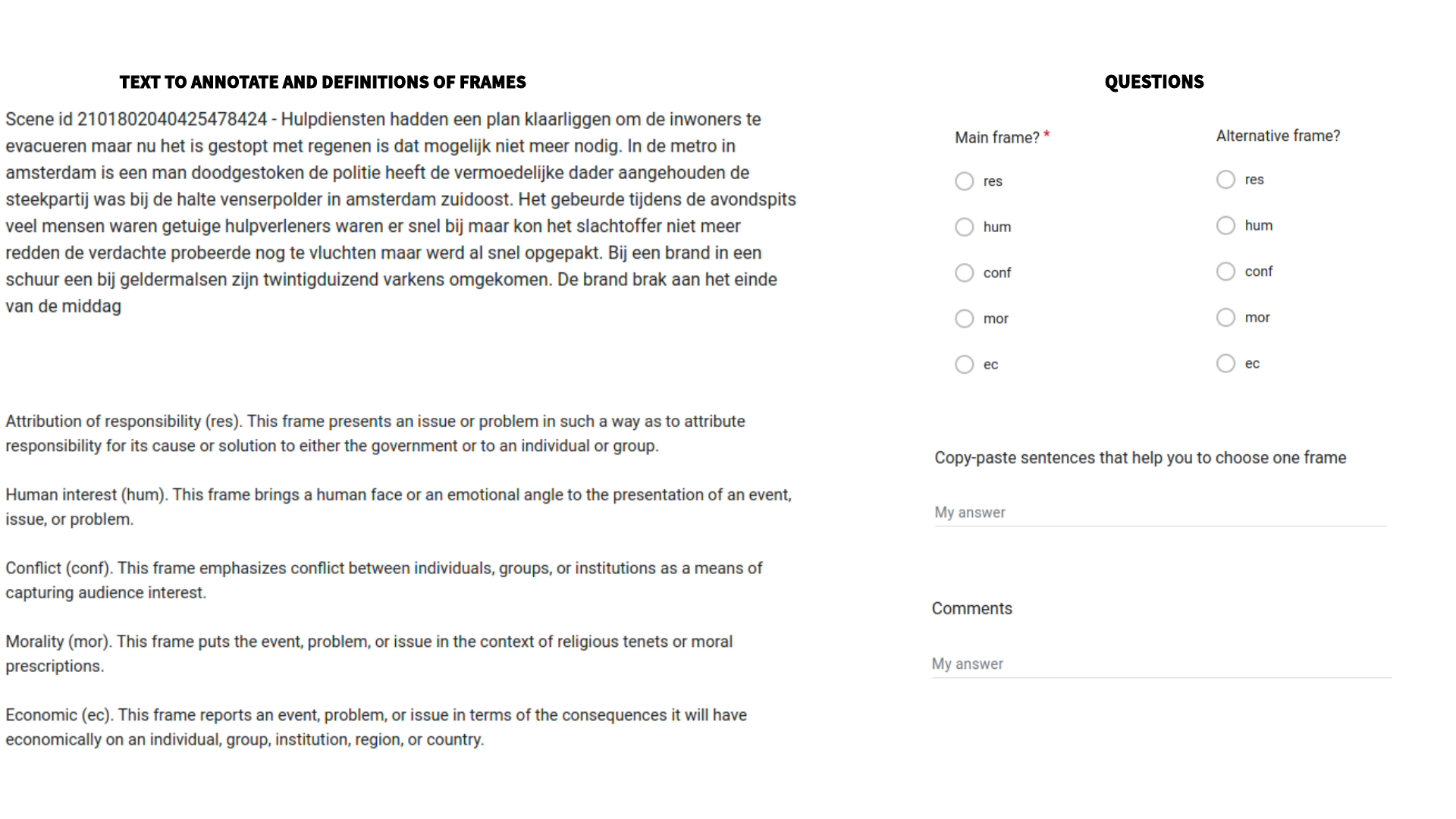}
  \caption{Example of the structure (text to annotate + definition of frames + questions ) of the form to do the annotation.}
  \label{fig:form}
\end{figure}

After these definitions, we asked the annotator (1) to define the main frame; (2) to define an alternative frame if there was one; (3) to copy-paste sentences that helped the annotator chose the main frame; and (4) to add free text in a section called comments, in case that the annotator had something to explain. Figure \ref{fig:form} shows an example of a form.

We created 20 forms for each program, where each form had 50 pieces of texts to annotate. Subsequently, the answers to these forms were saved for later analysis.

\subsection{Classification with GPT-3.5}
\begin{figure}[h]
  \centering
\includegraphics[width=\linewidth]{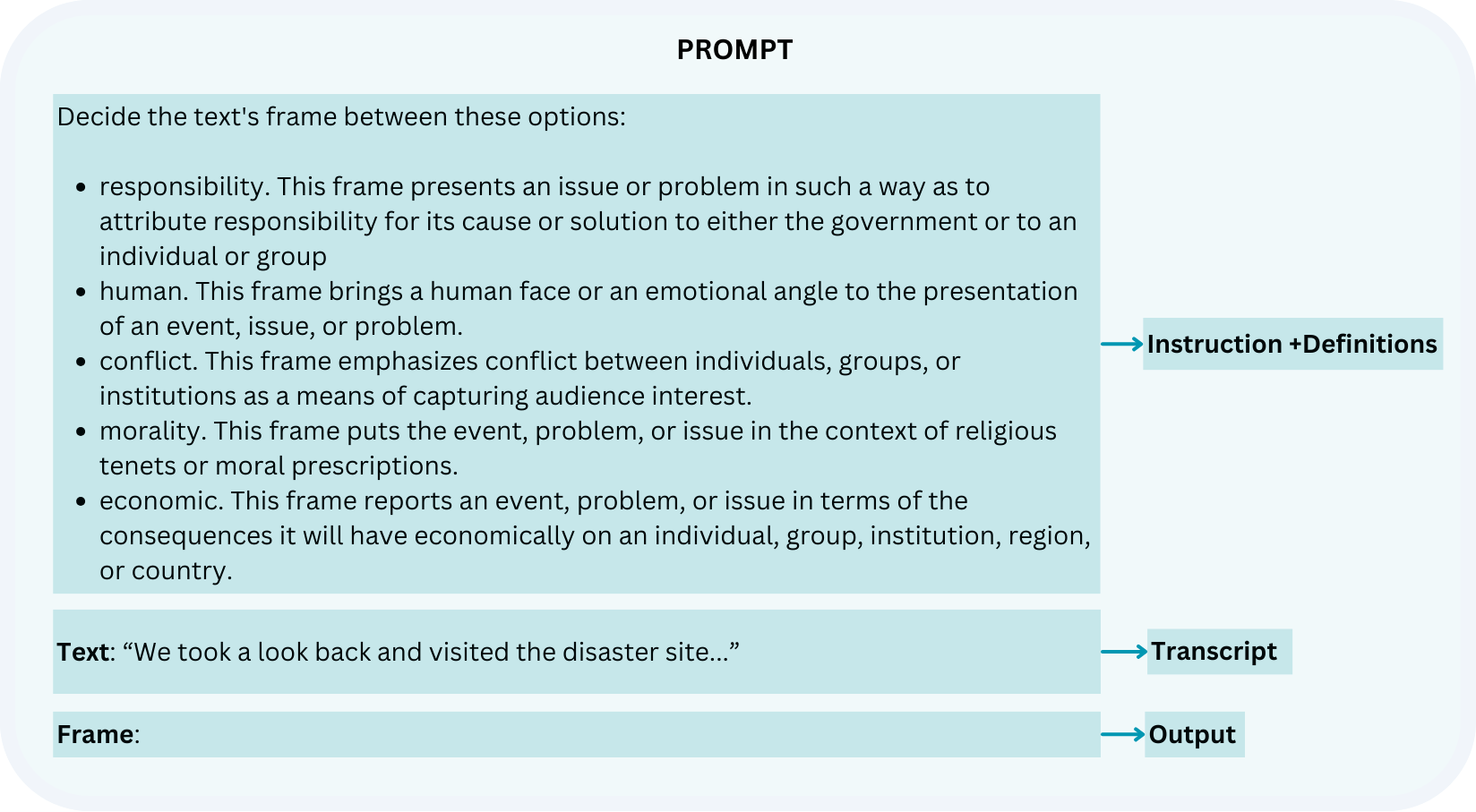}
  \caption{Example of the prompt used for frame classification, given a transcript.}
  \label{fig:schema_prompt}
\end{figure}

The generative language model is used as follows. This type of model, given an input that we pass (called prompt) is able to generate text that continues that prompt (called output.) As a simple example, if we ask a generative model "how many sides does a triangle has?", it will generate an output through a series of tokens (a token can be a word, or a smaller unit, so a word can be formed by more than one token), and those tokens have a probability, which reflects how confident the model is of the answer, based on the text it has been trained with. In this simple example, the answer would be "three" with a 100\% probability of that token. Based on this idea (i.e., that the generative model produces an answer and gives that answer a probability), we defined a prompt in which we first pass the definitions of the different frame types, then we pass the text to classify, and finally we ask the model, among the 5 frame type options that we gave, which was the most likely frame. Figure \ref{fig:schema_prompt} shows an example of the prompt used. The model gives a probability to each of the 5 frame options, so the frame with the highest probability is the predominant frame identified by the model. Furthermore, the fact of being able to access the probabilities given to the other frame types, allows us to study cases where more than one frame was possible, because the second or third options had a high probability.
% \begin{figure}[h]
%   \centering
% \includegraphics[width=\linewidth]{images/_prompt.drawio.png}
%   \caption{Schema prompt passed to GPT-3.5.}
%   \label{fig:schema_prompt}
% \end{figure}

To use GPT-3.5 as a frame classifier, we use the OpenAI API \footnote{https://platform.openai.com/docs/introduction/overview}. As we mentioned before, 
% i eliminated this sentence as we just said the same in the previous paragraph
%what we passed as a prompt was the definitions of the 5 frame types, and then we passed the text to classify, and asked which was the most likely frame. As a result 
the model outputs the most probable token, which in this case corresponded to one of the frames, and also outputs the next most probable tokens, which correspond to other frames. In this way, we could save the probability associated with each type of frame for each text, being the frame with the highest probability the one that GPT-3.5 considered most predominant.

GPT3 has different parameters that adjust the randomness and creativity of the answer. We set the temperature to 0, since the higher the temperature the more random the answer. Moreover, the Top-p parameter was set to 1 (as it would likely get a set of the most likely words for the model to choose from). As a model, we used the one with the best performance at the time of experimental design, which was TEXT-DAVINCI-003. Once we defined the parameters and the prompt to use, we made a call to the API for each of the pieces of text, obtaining the most likely frame associated with each text and the probabilities associated with each frame, for further analysis.

\subsection{Understanding human-machine disagreement}
Once the human annotation and machine inference were produced as described earlier in this section, we evaluated the results by quantifying the cases of  agreement and disagreement between human and machine. In addition, we had access to the probabilities given by GPT3.5 to the label given by the human, which allowed us to analyze how high the probabilities were in the cases of agreement between the two, and at the same time how high the probabilities given by the machine to the human label were in cases of disagreement.

This analysis was done with the support of the annotator, as well as academic experts, through semi-structured interviews. This qualitative analysis enriches our understanding of how LLMs are used in this task.

% we could study how high was the probability in the cases of agreement, but also study what was the probability given by the machine to the human label in the cases of disagreement, to see if the machine detected that frame identified by the human or not.

More specifically, evaluation sessions with three framing analysis experts were held to obtain their input on the results of the research, and on the best possible ways to utilize the outcomes for future research. We held three evaluation sessions with 2 Full Professors and one Assistant Professor. We invited these three scholars due to their knowledge on framing analysis, computational methods, media and television heritage materials, and the use of audiovisual datasets. The interviews were held in June and July 2023. The interviews were semi-structured, and were centered around getting feedback on our research setup and results. To elicit responses, we first shared the project goals and the outcomes of the first phase of the research. 
In addition, the annotator was interviewed about her personal experiences labeling the items.  The interviews with experts and annotator were not recorded, but notes were taken. Additional input was provided by email.

\section{Results and Discussion} \label{section:results}

%In this section, we describe the performance of GPT-3.5 for both TV programs.

\subsection{Frame labeling: agreement between human and machine (RQ1)}

\subsubsection{EENVANDAAG}
\begin{figure}[h]
  \centering
\includegraphics[scale=0.5]{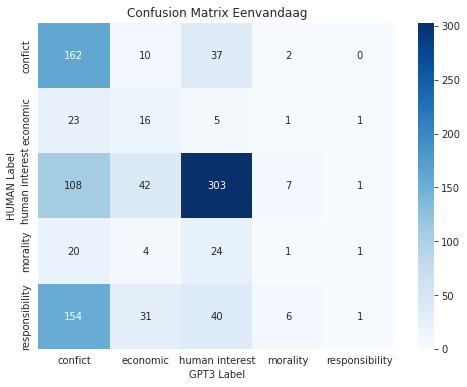}
  \caption{Agreement between human annotator and GPT-3.5 on classification of EenVandaag transcripts into five categories: Conflict, Economic, Human Interest, Morality and Responsibility.}
  \label{fig:e_confused}
\end{figure}

In the case of EenVandaag, the agreement between annotator and GPT-3.5 is 483 of 1000 items. This corresponds to an accuracy of 48.3\%. Of those cases of agreement, 303 are human interest, 162 are conflict, 16 are economic,  1 is morality and 1 is attribution responsibility. The confusion matrix is shown in Figure \ref{fig:e_confused}. We see that:

% In the figure \ref{fig:e_main_frame_human_gpt} we can see the distribution of main frames identified by the human and the GPT-3.5. We see that the predominant frame identified by the LLM is conflict mean while the annotator said that in most of the cases the frame is human interest.

% \begin{figure}[h]
%   \centering
% \includegraphics[width=\linewidth]{images/eenvandaag_machine_human_mainframe.png}
%   \caption{Main frame detected by GPT-3.5 and main frame detected by human in EenVandaag program.}
%   \label{fig:e_main_frame_human_gpt}

% \end{figure}

%We have compared the label given by the annotator versus the label given by the machine and see what happens in each type of frame trough a confusion matrix. The figure \ref{fig:e_confused} shows these results

\begin{itemize}

    \item Conflict (211) is identified by the machine in 162 cases. Then is confused with human interest (37) and economic (10).
    \item Economic label (46) is well detected in 16 items, and it is confused with conflict (23), human interest (5), morality (1), and attribution of responsibility (1).
    \item Human interest (461) is correctly identified by the machine in 303 cases, and it is confused with conflict (108), economic (42), morality (7) and attribution of responsibility (1).
    \item Morality (50) is not identified at all. The machine inferred human interest (24) and conflict (20) in most cases.
    \item Responsibility (231) is not identified at all. The machine most commonly inferred conflict (154), human interest (40) and economic (31).

\end{itemize}

\subsubsection{NIEUWSUUR}
\begin{figure}[h]
  \centering
\includegraphics[scale=0.5]{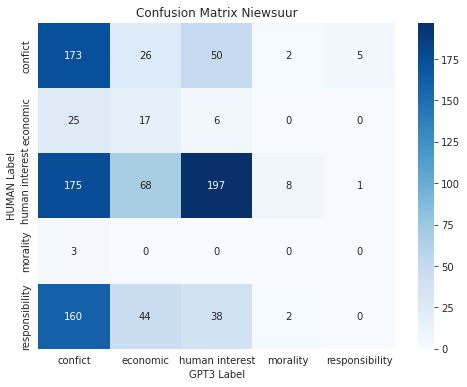}
  \caption{Agreement between human annotator and GPT-3.5 on classification of Niewwsuur transcripts into five categories: Conflict, Economic, Human Interest, Morality and Responsibility.}
  \label{fig:n_confused}

\end{figure}

In the case of Niewwsuur, the agreement between annotator and GPT-3.5 is 387 of 1000 cases. This corresponds to an accuracy of 38.7\%. Of these 387 cases, 197 are classified as human interest, 173 as conflict, and 17 as economic. Figure \ref{fig:n_confused} shows the confusion matrix. We observe that:
% In the Figure \ref{fig:n_main_frame_human_gpt} as before, we can see the distribution of main frames identified by the human and the GPT-3.5, but now for the items of Niewsuur. We see that more than half of the cases are labeled as conflict by GPT-3.5 mean while the annotator said that in most of the cases the frame is human interest.

% \begin{figure}[h]
%   \centering
% \includegraphics[width=\linewidth]{images/nieuwsuur_machine_human_mainframe.png}
%   \caption{Main frame detected by GPT-3.5 and main frame detected by human in Niewsuur items.}
%   \label{fig:n_main_frame_human_gpt}

% \end{figure}

% As before, we wanted to compare the label given by the annotator versus the label given by the machine and see what happens in each type of frame. 

\begin{itemize}
    \item Conflict (256) is identified by the machine in 173 cases. Then is confused with human interest (50), economic (26), attribution of responsibility (5) and morality (2).
    \item Economic label (48) is identified in 17 cases, and is confused with conflict (25), human interest (6).
    \item Human interest  (449) is identified by the machine in 197 cases. Confused with conflict (175), economic(68), and attribution of responsibility (1).
    \item Morality (3) is not identified at all. The machine inferred conflict in all the cases(3).
    \item Attribution of responsibility (244) is not identified at all. The machine most often inferred  conflict(160), human interest (38), economic (44) and morality (2).
    
\end{itemize}

\subsection{Effect of text length (RQ1)}
We investigated whether text length had an influence on the cases of agreement and disagreement between human and machine. We hypothesized that longer texts might be more likely to have more than one frame, and that this may in turn lead to disagreement, while shorter texts might be more likely to have only one frame.

It is important to consider the differences in the number of articles between the two categories (human-machine agreement and human-machine disagreement ) when comparing their distributions of article lengths, 
because the differences in sample sizes can affect the results. For example, in our case we have more data points of disagreement and the percentages of disagreement would be higher simply due to the larger number of data points, which could give a misleading impression of the data distribution

To limit this issue, we normalized the distributions by calculating relative frequencies. Instead of looking at the absolute number of articles in each length category, we have calculate the percentage of articles within each length category for both agreement and disagreement. This allows us to compare the distributions while accounting for the differences in sample sizes. Similarly, the length of the articles is not equally distributed, therefore we have divided the articles into bins based on word count. We have created bins of 100-199 words, 200-299 words, and so on, up to 800 words. This allowed as to categorize articles into different word count ranges. Within each word count range, we counted the number of cases of disagreement between human and machine and the number of cases of agreement. Later we have normalized the counts within each word count range by dividing the counts by the total number of articles in that range. This gave us the percentage of cases of disagreement and agreement within each word count range.
\subsubsection{EENVANDAAG}
\begin{figure}
  \centering
  \begin{subfigure}{0.45\textwidth}
    \centering
    \includegraphics[width=\linewidth]{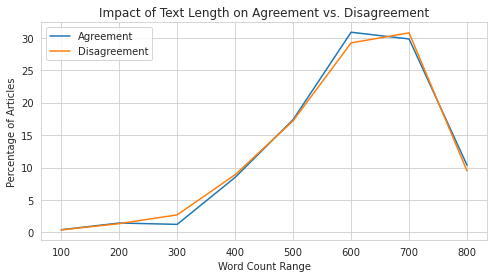}
    \caption{EenVandaag}
    \label{fig:e_length}
  \end{subfigure}
  \hfill
  \begin{subfigure}{0.45\textwidth}
    \centering
    \includegraphics[width=\linewidth]{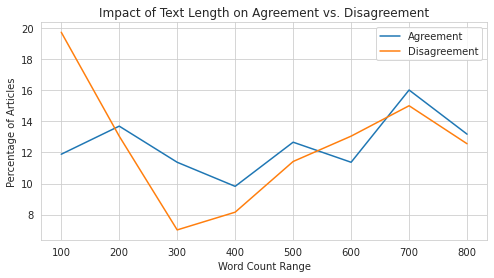}
    \caption{Nieuwsuur}
    \label{fig:n_length}
  \end{subfigure}
  \caption{Relative frequencies of agreement and disagreement between human and GPT-3.5 annotations for EenVandaag/Nieuwsuur transcripts across text length bins.}
  \label{fig:relative_}
\end{figure}

In Figure \ref{fig:e_length}, the x-axis represents the word count ranges (e.g., 100-199, 200-299, etc.), and the y-axis represents the percentage of cases of disagreement and agreement. 
% As we can see, the distributions follow a very similar tendency. In the range between 100 and 700 words the more words, the more cases of agreement, but also the more cases of disagreement. In the case of 800 words we see that this tendency of more words, more agreement and more disagreement is not satisfied, but in any case, the distributions are very similar. The length of the texts does not seem to be a factor that influences the performance of GPT-3.5. 
As we can see, the figure reveals similar patterns for agreement and disagreement across text lengths. Between 100 and 700 words, there's a general increase in both agreement and disagreement with more words. However, the 800-word bin deviates from this growth trend, but in both cases agreement and disagreement are declining in a similar way (in the last bin there were not many samples). Overall, these findings suggest that text length might not be a significant factor influencing GPT-3.5's performance.

In addition, as we have cases with more than one frame (the annotator had the option to choose an alternative frame to the first choice in cases of doubt), we wanted to check  if long texts are more likely to have more than one frame (alternative frame), or if, on the contrary, a longer text defines more clearly a single frame (non alternative frame). 

\begin{figure}
  \centering
  \begin{subfigure}{0.45\textwidth}
    \centering
    \includegraphics[width=\linewidth]{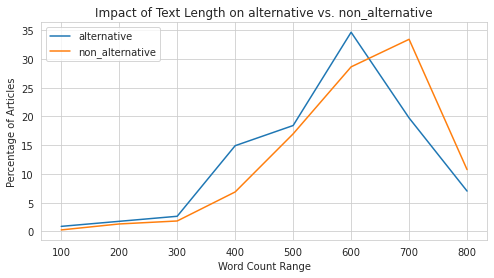}
    \caption{EenVandaag}
    \label{fig:e_length_alternative}
  \end{subfigure}
  \hfill
  \begin{subfigure}{0.45\textwidth}
    \centering
    \includegraphics[width=\linewidth]{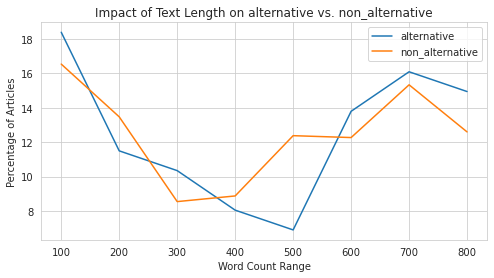}
    \caption{Nieuwsuur}
    \label{fig:n_length_alternative}
  \end{subfigure}
  \caption {Relative frequencies of alternative/non alternative frame between human and GPT-3.5 annotations for EenVandaag/Nieuwsuur transcripts across text length bins.}
  \label{fig:length_comparison}
\end{figure}

In Figure \ref{fig:e_length_alternative},  we can observe the percentage of number of words in 2 cases, in blue the cases with an alternative frame, and in orange the cases without an alternative frame. Both lines follow a similar pattern, remaining close to each other across most word count ranges. This indicates that text length has minimal impact on the presence of alternative frames. There might be slight variations between the lines in some bins, but they are not consistent enough to suggest a significant relationship.
% \begin{figure}[h]
%   \centering
% \includegraphics[scale=0.5]{images/e_length_alternative_frame_normalized_comparation.png}
%   \caption{Normalized distribution of number of words in the cases with more than one frame and the cases with only one frame detected by the annotator in Eevandaag program.}
%   \label{fig:e_length_alternative}
% \end{figure}

\subsubsection{NIEUWSUUR}
 In Figure \ref{fig:n_length} we see a slightly more noticeable difference than in the previous case between the cases of agreement and disagreement. In the case of 100 words bin, we see 8\% more of disagreement while in the case of 300 words there is around 10\% more of agreement articles. For the range between 400 and 800 words we see a fairly similar distribution.
% \begin{figure}[h]
%   \centering
% \includegraphics[scale=0.5]{images/n_length_texts_normalized_comparation.png}
%   \caption{Normalized distribution of number of words in the cases of agreement and disagreement in Nieuwsuur program.}
%   \label{fig:n_length}
% \end{figure}

In the case of the relation between alternative frame and length of the piece of texts, as it happened in the previous case, we do not appreciate any correlation between them analyzing the results in Figure \ref{fig:n_length_alternative}, because the trends are quite similar. We only see a relevant difference between the articles with and without alternative frame of 10\%  in the 500 words bin.
% \begin{figure}[h]
%   \centering
% \includegraphics[scale=0.5]{images/n_length_alternative_frame_normalized_comparation.png}
%   \caption{Noramlized distribution of number of words in the cases with more than one frame and the cases with only one frame detected by the annotator in Nieuwsuur program.}
%   \label{fig:n_length_alternative}
% \end{figure}

\subsection{Analyzing GPT output probabilities (RQ1)}
\subsubsection{EENVANDAAG}
\begin{figure}[h]
  \centering
\includegraphics[width=\linewidth]{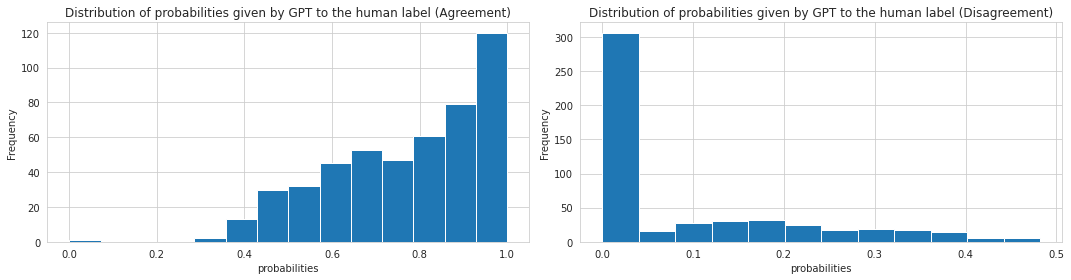}
  \caption{Distribution of the probabilities given by GPT-3.5 to the annotator label in the cases of agreement and disagreement.}
  \label{fig:e_probabilities}
\end{figure}

In Figure \ref{fig:e_probabilities}, we can study the distribution of probabilities given by GPT-3.5 to the human label, in the cases of agreement and disagreement.

In the case of human-machine agreement, we see that the probabilities are high, more than 50\% in most cases, mean while in the case of disagreement, we see that the probabilities are very low, practically null i.e., the machine does not detect the frame chosen by the human. 

\subsubsection{NIEUWSUUR}
\begin{figure}[h]
  \centering
\includegraphics[width=\linewidth]{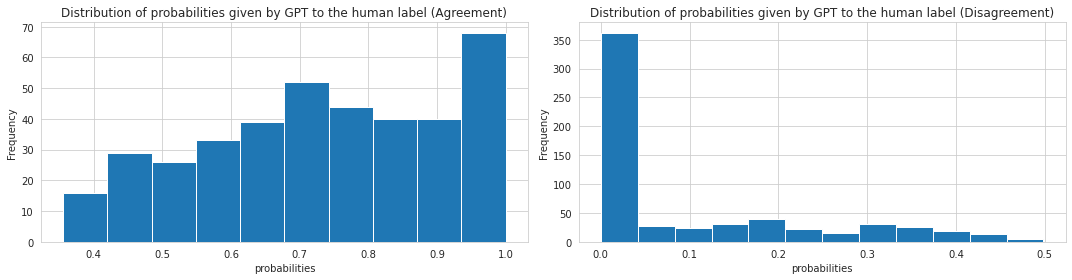}
  \caption{Distribution of the probabilities given by GPT-3.5 to the annotator label in the cases of agreement and disagreement.}
  \label{fig:n_probabilities}
\end{figure}

Regarding the probabilities given by the machine to the human label in Figure \ref{fig:n_probabilities}, as it happened previously we observe a high probability in the cases of agreement, but really low in most of the cases of disagreement, that means that the human level is not identified at all in many cases by the GPT-3.5. In the right part of the figure we see how the probability associated with the human label in the cases of agreement is higher than 40\% in most cases, while in the figure on the left we see how the probability given by the machine to the human label in the cases of disagreement is zero in most cases, so there are two distinct cases, cases where the frame is detected with a high probability, because having more than 40\% probability in the choice of the main frame, having 5 possible options is quite high, but at the same time there are many cases where the frame has not been identified by the machine at all giving probabilities of zero.
\subsection{Contextualizing the results: Comparison with related work (RQ1)}

In this section, we put in context the results obtained, making a comparison with previous literature.

Research on identifying frames through the use of LLM is in an incipient phase, and emerging literature is gradually addressing this issue by using different types of texts to be classified, as well as different models or number of labels in the classification. As a result, we cannot make a direct comparison, since to the best of our knowledge, this is the first work where frames are automatically identified in TV news show transcripts. In any case, an indirect comparison helps to better understand the complexity of the task and the different approaches used to tackle it. Table \ref{tab: related_work} presents a summary of the main characteristics of each work.

In the work of Gilardi et al. \cite{Gilardi_2023} they classified the frames of tweets with a system of 15 different types of frames using ChatGPT. They work with tweets and a frame classification system of 15 different types, although in this case they use ChatGPT as a model with different parameter settings. In a dataset of tweets from 2020 and 2021, the accuracy of the model is around 40\%, while a dataset of tweets from 2017-2022 is a little over 50\%. Despite using another model, a system with a higher number of tags, and different type of media content, the human-machine agreement is very similar to our case.

\begin{table}[!htp]\centering
\caption{Summary of related work using LLMs for frame classification.}\label{tab: related_work}
\scriptsize
\resizebox{8.5cm}{!}{
\begin{tabular}{lccccc}\toprule
\textbf{Related work} &\textbf{Type of data (avg.\# words)} &\textbf{\# labels} &\textbf{LLM} &\textbf{acc(\%)} \\\midrule
Gilardi et al.\cite{Gilardi_2023} &tweets (30) &15 &ChatGPT &40-50 \\
Alizadeh et al. \cite{alizadeh2023opensource} &tweets (30) &15 &HuggingChat \& FLAN &30-40 \\
Alonso del Barrio et al.\cite{alonsoframing} &headlines (8) &6 &GPT3.5 &49 \\
% Hoes &claims (40) &6 &ChatGPT &29 \\
Ours &transcripts (600) &5 &GPT3.5 &43 \\
\bottomrule
\end{tabular}
}
\end{table}
In the work of Alizadeh et al. \cite{alizadeh2023opensource}, which is related to the one mentioned above (the same data and the same frame classification system, but this time open-source models are used),  the performance of open-source for that classification system is a bit worse than ChatGPT, yet in other annotation tasks the open-source version seems to work better.

In our previous work \cite{alonsoframing}, we followed a  similar approach, with two main differences, on one hand we used headlines from newspaper articles as text to classify, and on the other hand we used the same types of frames as we do here, adding an extra type of frame (called no-frame), since for such short sentences, there could be cases where indeed there was no frame. The results showed a 49\% agreement between human and machine, which is not very far from our results, and reflects the complexity of the task as it entails significant subjectivity. In that work, we performed a post-hoc experiment where annotators were asked if they agreed or disagreed with the label provided by GPT-3.5 (without knowing the origin of this label), and they agreed in 75\% of the cases. 
It is this type of result that illustrates the potential that LLMs can have as an annotator support tool, maintaining human-in-the-loop validation during the annotation process to ensure the reliability and accuracy of the annotations \cite{thapa2023humans}.

Given the discussed results, we see that our results are within an acceptable range, given that we have used much longer texts than those used in previous literature, thus making the task more difficult. Nevertheless, we still see potential and much room for improvement in the field of frame analysis with LLMs.
\subsection{Expert and annotator feedback on human-machine agreeement (RQ2)}
{\bf General remarks.} Both the scholars and the annotator concluded that the human labeling process is time-consuming and exhausting, and thus influences the results over time. Whereas the computational labeling remains consistent over time, the human labeling does not. Because of this, the annotator stated that over time, she was more likely to choose labels that were better known to her, because those were the labels that she used most.

Due to limited time and resources, we  worked with a single annotator. Although she received an extensive briefing on how the labels should be attributed, the results are nevertheless influenced by personal preferences and experiences. As all scholars pointed out, for this research to become more valuable and provide more factual information, there is a need for multiple human annotators. This information accounts for the fact that some labels were chosen more often than others. Furthermore, in cases the annotator was uncertain about a specific label, she added an alternative label. %Because of this, we can conclude that in many cases, it was unclear which label worked best. 
All these cases should, in future research, be analyzed in relation to other human annotators.

Finally, the automatic transcription of the labels proved hard to read for the annotator. Some of the articles were scrambled because the spoken texts in the news item were unclear. %This resulted in uncertainty about certain labels, especially with the conflict label. It was not always clear whether there was conflict in the framing of the item. Often, the annotator would add an alternative frame to the text because she was unable to choose.

\textbf{Experts' in-depth views.} In response to the examination of the results of agreement/disagreement between human and machine, the experts discussed their hypotheses regarding the disagreement of labels. One of them suggested that "\textit{the confusion between human interest and attribution of responsibility with the label conflict might be due to the intertwined nature of these frames}". The expert also argued that these frames are essentially building blocks of the same narrative, as a conflict often entails identifying responsibility and putting a face on an issue. The expert also emphasized that the distinction between conflict, responsibility, and human interest is based on the taken perspective, i.e., "\textit{with conflict focusing on format, responsibility on actors in the conflict, and human interest involving one of these actors}".

%JASMINE
Another expert provided two potential reasons for the disagreement. First, it was noted that human interest and attribution of responsibility often appear together in news stories, as "\textit{responsibility is more sentiment-based, while human interest revolves around thematic elements}. This co-occurrence could lead to confusion. Second, the expert suggested that "\textit{the framing cues used in the input data might not be well-defined and clear-cut enough}". The lack of precise distinctions between these frames may result in ambiguity, causing GPT-3.5 to mix them up during inference.
%interpretation.

In summary, both experts offered insights into why GPT-3.5 might confuse human interest and attribution of responsibility with the label conflict. They highlighted the interconnected nature of these frames and the potential co-occurrence of elements in news stories as contributing factors to the disagreement. Additionally, they mentioned the importance of  clear framing cues to avoid such confusion.

%BENTHE
{\bf Annotator's in-depth views.} In the case of the annotator, she suggested that the disagreement may be due to nuances in the content, emphasizing that these nuances are often quite subtle. For instance, when distinguishing responsibility from conflict, the annotator typically considered "\textit{situations where one party is accusing another of causing a problem or demanding government intervention to address an issue. In their view, these distinctions might be too nuanced for a machine to grasp. The AI might simplify the situation and classify it simply as a conflict between two parties, even if one party is claiming responsibility.}"

Regarding human interest, the annotator explained that this frame was typically chosen when the content focused on individual cases within a broader context. She provided an example of the refugee crisis, where she would select the human interest frame when the narrative zooms in on a specific refugee, like a woman in a refugee camp, and explores her life. However, she acknowledged that "\textit{there is a larger conflict underlying these human interest stories because people wouldn't become refugees if there were no conflicts}". The annotator suggested that "\textit{these subtleties might not be readily detectable by the machine, leading to the misclassification of frames."}

In summary, the annotator's views were centered on the idea that the nuances within the content might be challenging for the machine to distinguish accurately, resulting in the misclassification of frames like responsibility, conflict, and human interest. 

\subsection{Expanding and updating framing categories (RQ2) }
We also asked the experts and the annotator if they could provide with alternatives frames they considered appropriate. The the answers were very varied, and are summarized below.

%ALTERNATIVE BALDWING
{\bf Experts' views.} One of the experts pointed out to be supportive 
%a 'cool lover' 
of the typology by Semetko and Valkenburg. "\textit{Their view was that there was not a limited set of frames that journalists habitually choose from, which they could then apply to all possible topics}. While acknowledging the existence of these five frames within the studied set, the expert believed that "\textit{the frames chosen in newsrooms were more issue-specific and played a vital role in giving meaning to social issues.}" \cite{semetko}.

In the expert's opinion, when thinking about a more generic frame that could apply to many issues, they suggested something like the "\textit{politics-are-responsible frame}." This frame focused on "\textit{how administrators, governments, and politicians are responsible for many issues, either as a cause or as a solution}. Additionally, the expert  noted that many social issues were often defined in terms of "\textit{left-wing and right-wing perspectives, which went beyond the conflict frame}". The expert saw that the conflict frame was as omnipresent as it was elusive, emphasizing that it would be more relevant to consider whether left or right forces were being examined in the process.

Furthermore, the expert identified the '\textit{system frame}' as relevant, highlighting the existence of '\textit{structures}' responsible for various issues. The expert viewed the '\textit{system}' as something less tangible and partly non-human, citing examples like the invisible hand in economics. The expert believed that this frame was often used to evade individual responsibilities, such as during the banking crisis when banks referred to a systemic problem to deflect blame. The expert raised questions to explain this concept: "{\textit who controls the system?, how is it maintained?, and what do the underlying structures look like?}

Additionally, the expert mentioned the concept of "\textit{the people}" as a frame, where a vague notion of public opinion or desires was referenced without further scrutiny. The expert highlighted that it was important to recognize that "\textit{the people are not a entity}" and that various perspectives existed within the population.

% Alternative JASMINE
Another expert offered an alternative perspective on the frames used in the experiment, particularly the conflict, human interest, economic, morality, and attribution of responsibility frames from the Semetko typology. This expert suggested the possibility of incorporating more positive frames, such as "\textit{reconciliation as an alternative to conflict.}" They also proposed rephrasing human interest to a more "\textit{sentimental frame, like vulnerability, to align it better with the sentimental frames of morality and responsibility."}

The expert noted that \textit{"human interest," "economic," and "conflict" could be seen as overarching themes of a broadcast or news story, while "morality" and "responsibility" were more implicit or sentimental in nature}. This suggested that the first three frames could be easily distinguished from each other, while the last two may be more challenging to differentiate.

%In summary, the expert proposed considering positive frames like "reconciliation" and rephrasing "human interest" to a more 'sentimental' frame like 'vulnerability' for a more cohesive framing approach. The expert also commented on the distinctiveness of the chosen frames, with the first three being more explicit, and the last two being more implicit or sentimental in nature.

% BENTHE
{\bf Annotator's views.} As for the annotator, she expressed that the frame of morality was the one she used the least and the one she found most difficult to identify, perhaps because of its more religiously oriented definition. As for the economic frame, she did not use it much either, but it was an easy frame to identify. She commented that "\textit{there were cases where the news apparently did not seem to have any frame, but was very neutral}" and perhaps an alternative frame that was "no frame" would have been useful, because sometimes she was forced to choose at least one, when, in her opinion, none of them fitted correctly in reality.

%\subsection{Comparison with related work and a look at its potential application (RQ3)}

\subsection{Moving forward: from current work to future applications (RQ3)}

Regarding issues to be improved in LLM classification, there are several paths to explore. 
First, at prompt level, we wanted to start with a zero-shot learning approach; we consider it the logical initial step, since there is no need for annotated data. Providing some examples (few shot learning) could likely improve the classification of frames; at the same time, we recognize the subjectivity involved in this process, and that there will likely be cases where more than one option is correct.

Furthermore, we are aware of the lack of transparency of closed-source models, yet the fast development of these LLMs allowed us to use them and try to identify their strengths and limitations. An open-source model trained directly on Dutch would be ideal, but the reality is that large open-source models (which can be compared to OpenAI models) require computational resources that few people can access. Based on these points, and understanding the reality of the situation and the tools available to us (at the time of the experiment, GPT3.5 was the most stable model on the market, since ChatGPT was a conversational model, not an instruction-based model, and GPT4 had just been released), we decided to use GPT3.5. This paper shows both the limitations and the potential that  LLMs have, as a complementary tool, to facilitate the work of media professionals in different scenarios.

% FUTURE APPLCIATIONS AND FUTURE RESEARCH
This type of models could even be used to, once an article has been generated by a journalist, ask which is the frame identified by the LLM with arguments. A journalist can potentially take some advantage of the bias of the models, because these models have a bias according to the text with which they have been trained \cite{grossmann2023ai, feng2023pretraining}, but audiences also have bias, so the model can reflect part of the audience. We believe it is a useful experience to do this type of exercise, and that the two options are valid: if there was agreement between human and machine, perhaps to show that the frame is very evident; if there was disagreement, it could be suggested that the article may have more than one valid frame, as long as the arguments given by the machine are convincing. 

In other recent research, the work of Petridis et al. \cite{maiden2018making} is an example of an interactive tool for journalists using LLMs to explore angles for reporting a press release \cite{petridis2023anglekindling}. This work is an improvement of the performance of a previous work that uses traditional natural language processing techniques instead of LLMs \cite{maiden2018making}

Looking forward to the future use of LLMs in media, the work of Naoain et al. \cite{naoain2022addressing} argued that a change of mindset in the media ecosystem is needed, and training on the use of these AI tools must be a priority given the lack of existing knowledge.

% For media students:
Media students need to develop an understanding of framing theory in order to critically analyze media messages. One potential direction of our work is the creation of interactive learning modules that allow students to practice identifying frames in media content or to explore the different ways that framing can influence audience perceptions. A potential application we can see is that just as  annotators read a text and annotated the frame they considered giving arguments as to why they chose that frame, the frame chosen by the LLM could be shown with the machine's "arguments". In this way, there could be cases where the perception is the same but there could also be cases where it is different, and thus elicit the realisation that there are many ways of thinking and that one personal perception in many cases does not mean that it is the only possible valid one. 

We also envision that the use of these tools can benefit and facilitate the work of students, and there is a lot of research being done on how LLMs can be used in education \cite{filgueiras2023artificial}.
Gan et al. \cite{gan2023large} define practical areas of the use of LLMs in Education such as instructional support tools, as assistants to teachers, providing intelligent instructional support tools and platforms, and  educational assessment and feedback with learning data to provide assessment and feedback on their learning progress.

In summary, we believe that the future directions outlined here are logical implications of the work presented in this paper, and clearly need to be deepened and tested as part of future work.

\section{Conclusions}\label{section:conclusions}
In this article, we study the use of prompt-engineering for automatic frame inference in TV content. We conclude by answering the three research questions we posed:

RQ1:  How accurately can a GPT model identify frames in television programs using a prompt-engineering approach?
Taking advantage of text generation models for prompt-engineering classification tasks is a very promising option, as no annotated data are needed for their training phase. There is much room for improvement, both in the definition of the prompt and in the performance of such models but in any case, we see potential in the use of this tool to support journalists.

RQ2:  In comparison to human annotation, what are the limitations of a GPT-3.5 model in classifying frames of TV show transcripts?
We have seen that we have obtained results below 50\% agreement between human and machine, but this is within the results obtained to date in previous literature. We also consider that the quality of the transcripts has influenced the understanding of both the annotator and the LLMs, as well as the possibility that, being a task involving some subjectivity, there is more than one correct answer and in future work we must have more annotators, as well as explore new types of frames that deal with the subtleties of language that can provoke human-machine disagreement.

RQ3: How can GPT models and prompt engineering be used in potential interactive applications of framing analysis?
We have discussed the potential uses of this type of tools in the professional environment where there is a strong tendency to the immersion of information technology in the newsrooms, but also from the academic point of view, where the professionals of tomorrow in the journalistic world have to adapt to technological advances.

% GPT-3.5 for the study cases has not worked well, as there are a lot of cases where the human said a frame and the machine did not even give probability to that frame. It is true that the data has been annotated by only one person, and as a possible continuation of the research several avenues are open. One of them would be to do an annotation a posteriori, and see if the annotator agrees with the frame given by GPT-3.5, because it may happen that there is more than one possible option. Another possible path is a deductive method. In this method, we would add multiple human coders to get a broader corpus of interpretations of the data. We would not only ask them to label the data but also provide us insights in their motivations. We will then feed these motivations back to the LLM by using prompt engineering. That way we would try to get a closer match of the artificial framing to the human framing so that we can use the artificial framing for future framing analysis. This would hypothetically save future scholars a lot of time and resources since they will be able to work with less human coders. The deductive method is the most promising in the eyes of the scholars.  There is also the possibility to test other LLMs that have been appearing last days. We could also consider an Inductive method. In this method, we would come up with a new set of frames that is specific for the corpus of news items that we analyze. 

\begin{acks}
We thank Prof. Baldwin van Gorp, Dr. Jasmijn van Gorp and the third expert for their contributions to the analysis of the results. We also thank Benthe van Hofwegen for her annotation work and contributions to the analysis of the results. This work was supported by the AI4Media project, funded by the European Commission (Grant 951911) under the H2020 Programme ICT-48-2020.
\end{acks}

\bibliographystyle{ACM-Reference-Format}
\bibliography{sample-base}

\end{document}